\documentclass[iop,apj,twocolumn]{emulateapj_pab}
\RequirePackage{hyperref}
\hypersetup{
	colorlinks = true,
	linkcolor = blue,
	citecolor = blue,
	filecolor = blue,
	urlcolor = blue,
	pdfborder = {0 0 0}
}

\usepackage{color}
\usepackage{bm}

\renewcommand{\vec}[1]{\bm{#1}}

\graphicspath{{./}{./Plots/}{./Plots/Model/}{./Plots/Helios/}{./Plots/Fitted/}{./Plots/Helios_Model/}}

\received{2017 September 22}
\revised{2018 January 17}
\accepted{2018 February 6}
\published{2018 March 13}

\submitted{}
\journalinfo{The Astrophysical Journal, {\rm 855:111 (10pp), 2018 March 10 \hfill \url{https://doi.org/10.3847/1538-4357/aaae04}}}
\shorttitle{Electromotive Force in a CME Shock Front Model versus {\em Helios} Observations}
\shortauthors{Bourdin, Hofer, \& Narita}

\begin{document}
%\linenumbers

\title{Inner Structure of CME Shock Fronts Revealed by the Electromotive Force and\\
Turbulent Transport Coefficients in {\em Helios-2} Observations}

%\author{Philippe-A. Bourdin}
%\affiliation{Space Research Institute, Austrian Academy of Sciences, Schmiedlstr. 6, 8042 Graz, Austria}
\author{Philippe-A.~Bourdin$^{1}$ \orcid{0000-0002-6793-601X}, Bernhard Hofer$^{1,2}$ \orcid{0000-0002-8628-7887}, Yasuhito Narita$^{1,2,3}$ \orcid{0000-0002-5332-8881}}
\affiliation{$^{1}$ Space Research Institute, Austrian Academy of Sciences, Schmiedlstr. 6, A-8042 Graz, Austria, \href{mailto:Philippe.Bourdin@oeaw.ac.at}{Philippe.Bourdin@oeaw.ac.at}}
\affiliation{$^{2}$ Institute of Physics, University of Graz, Universit{\"a}tsplatz 5, A-8010 Graz, Austria}
\affiliation{$^{3}$ Institut f{\"u}r Geophysik und extraterrestrische Physik, Technische Universit{\"a}t Braunschweig, Mendelsohnstr. 3, D-38106 Braunschweig, Germany}

%\correspondingauthor{Ph.-A.~Bourdin}
%\email{Philippe.Bourdin@oeaw.ac.at}

%%%%%%%%%%%%%%%%%%%%%%%%%%%%%%%%%%%%%%%%%%%%
% Abstract
\begin{abstract}
Electromotive force is an essential quantity in dynamo theory.
During a coronal mass ejection (CME), magnetic helicity gets decoupled from the Sun and advected into the heliosphere with the solar wind.
Eventually, a heliospheric magnetic transient event might pass by a spacecraft, such as the {\em Helios} space observatories.
Our aim is to investigate the electromotive force, the kinetic helicity effect ($\alpha$ term), the turbulent diffusion ($\beta$ term) and the cross-helicity effect ($\gamma$ term) in the inner heliosphere below 1\,\rm{au}.
We set up a one-dimensional model of the solar wind velocity and magnetic field for a hypothetic interplanetary CME.
Because turbulent structures within the solar wind evolve much slower than this structure needs to pass by the spacecraft, we use a reduced curl operator to compute the current density and vorticity.
We test our CME shock-front model against an observed magnetic transient that passes by the {\em Helios-2} spacecraft.
At the peak of the fluctuations in this event we find strongly enhanced $\alpha$, $\beta$ and $\gamma$ terms, as well as a strong peak in the total electromotive force.
Our method allows us to automatically identify magnetic transient events from any in-situ spacecraft observations that contain magnetic field and plasma velocity data of the solar wind.
\end{abstract}


%%%%%%%%%%%%%%%%%%%%%%%%%%%%%%%%%%%%%%%%%%%%
\section{Introduction} \label{introduction}

A much discussed topic in plasma physics and astrophysics is the understanding of the role of turbulence in the solar wind \citep{Bruno+Carbone:2013}.
In the turbulent dynamo theory, small-scale plasma motions induce small-scale magnetic fields and interact with them to generate and sustain a large-scale magnetic field through the turbulent electromotive force, making it a quantity of primary importance (see \cite{Brandenburg+Subramanian:2005} and works cited therein).
This electromotive force has been measured extensively under laboratory conditions, most prominently in the study of plasma turbulence in nuclear fusion \citep{Ji+Prager:2002}.
A principal approach is the reversed field pinch (RFP) model, which assumes a plasma confined in a toroidal system by a poloidal magnetic field induced by a toroidal current and toroidal magnetic field from an external source \citep{Bodin+Newton:1980}.
However, the study of the electromotive force is not limted to laboratory experiments.
Numerical simulations provide an excellent way of improving our understanding of turbulent flows.
The advantage of using a direct numerical simulation is to get information that would be unobtainaible under laboratory conditions, at the cost of constraints such as boundary conditions.
This allows us to study in great detail the individual effects that may contribute to the electromotive force, e.g. the $\alpha$ effect \citep{Brandenburg:2001} and the cross-helicity effect \citep{Sur+Brandenburg:2009}.

\cite{Marsch+Tu:1992} made direct measurements of the fluctuating magnetic and velocity fields, based on {\em Helios} observations from the inner heliosphere.
Their results suggest that the $\alpha$ effect in the solar wind is negligible.
We do not expect to reveal a strong dynamo action in a quiet solar wind.
However, during a magnetic transient event, there might be turbulent vortical plasma motions that could give rise to a turbulent electromotive force.
We investigate transition layers of magnetic transients such as the shock front of a coronal mass ejection (CME) that might have strong turbulent plasma motions that could feature observable signatures of the electromotive force.
These plasma motions might decay over time and therefore be best observable close to the Sun.
Therefore, in this work, we focus on magnetic transients in the inner heliosphere.

The goal of this study is to understand the internal magnetic structure of shock fronts caused by CMEs that propagate through the inner heliosphere up to a distance of 1 astronomical unit (\rm{au}) from the Sun.
We would like to determine whether there are vortical plasma motions or helical magnetic fields that may cause induction work and give rise to a turbulent electromotive force.
We aim to model and reproduce the observational signatures of shock fronts passing by the {\em Helios-2} spacecraft.

The derivation of the electromotive force in a turbulent system has to consider also the compressibility of the plasma.
This is in general also the case in the heliosphere.
However, when a CME expands into the heliosphere, its spatial scale grows linearily while the amplitude of the shock starts falling together with the solar wind velocity outside of roughly 0.4\,\rm{au}.
Once the CME decelerates, the shock front becomes quasi-static and less compressed the farther the CME propagates.
Therefore, we start our model without taking compressibility into account.

%%%%%%%%%%%%%%%%%%%%%
\subsection{Electromotive Force}

\begin{figure}
\begin{center}
\includegraphics[trim = 0mm 0mm 0mm 0mm, clip, width=8.8cm]{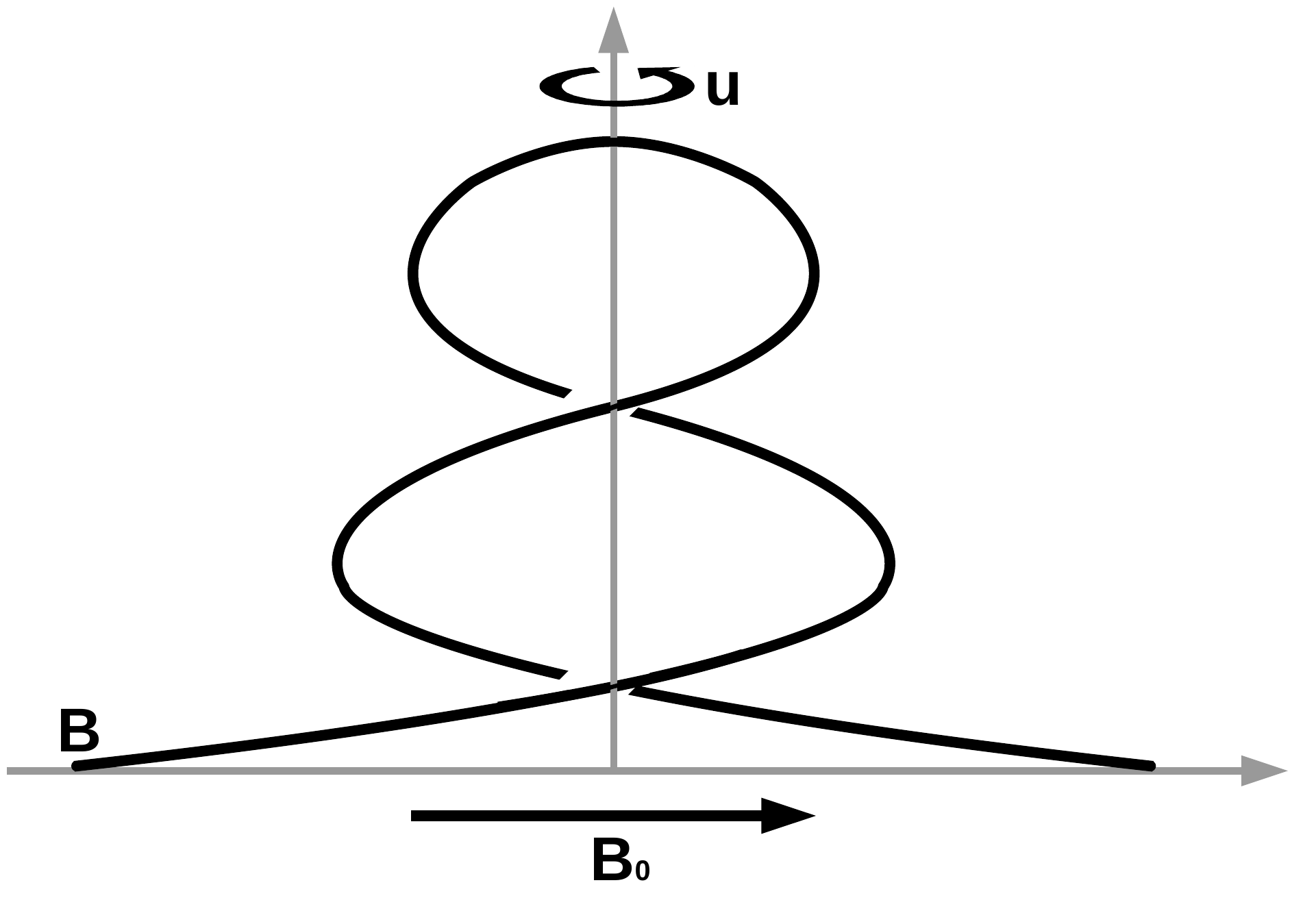}
\end{center}
\caption{Illustration of the $\alpha$ effect; see also chapter 9.3 in \cite{Priest:1982}.
$\vec{B}$ is the magnetic field, $\vec{u}$ the flow velocity.
A helical structure of the magnetic field is generated by a vortical plasma motion in a regime where the magnetic field is "frozen-in" into the solar wind plasma.}
\label{fig:vis_alpha}
\end{figure}

In mean-field electrodynamics, the magnetic field and the velocity field are split into two parts: the background field and the perturbation on this background; see \cite{Steenbeck+al:1966}.
The evolution of the background magnetic field $\vec{B}_0$ can be expressed by the mean induction equation:
\begin{equation}
\frac{\partial \vec{B}_0}{\partial t} = \vec{\nabla} \times (\vec{U}_0 \times \vec{B}_0) + \vec{\nabla} \times \vec{M} + \eta \vec{\nabla}^{2} \vec{B}_0,
\label{eq:ind}
\end{equation}
where $\vec{U}_0$ denotes the background flow velocity and $\eta$ is the magnetic diffusivity; see Eq.\,(22) in \cite{Yokoi:2013}.
The impact of small-scale turbulence on the large-scale magnetic field appears as an electromotive force $\vec{M}$.

The electromotive force can be derived from mean-field electrodynamics as the ensemble average of the cross product between the flow velocity fluctuations $\delta \vec{U}$ and the magnetic field fluctuations $\delta \vec{B}$:
\begin{equation}
\vec{M}_1 =  \langle \delta \vec{U} \times \delta \vec{B} \rangle.
\label{eq:em}
\end{equation}
However, Eq.\,(\ref{eq:em}) does not give information on the actual magnetic field generation mechanisms.

To circumvent the shortcomings of this formulation, there are alternative models for $\vec{M}$ that are based on theoretical or numerical dynamo studies.
Therefore, we also test a formulation based on the RFP model from \cite{Yoshizawa:1990} adjusted to SI units:
\begin{equation}
\vec{M}_{\rm{RFP}} = \alpha \vec{B}_0 - \beta \mu_0 \vec{J}_0 + \gamma \vec{\Omega}_0,
\label{eq:em2}
\end{equation}
where $\mu_0$ is the vacuum permeability.
\cite{Yokoi:2013} also uses this separation for helically driven magnetic turbulent dynamos.
The $\alpha$ term is proportional to the background magnetic field $\vec{B}_0$.
It represents the kinetic helicity effect and refers to the $\alpha$ dynamo mechanism.
Here, the large-scale magnetic field amplifies by turbulent helical twisting motions that convert poloidal into toroidal field, or vice versa.
A visualization of the $\alpha$ effect is given in Fig.\,\ref{fig:vis_alpha}; see \cite{Priest:1982}.
The $\beta$ term comprises a curl of the background magnetic field: $\mu_0\vec{J}_0 = \vec{\nabla} \times \vec{B}_0$.
It represents the turbulent diffusion effect ($\beta$ term) that causes diffusion of the large-scale magnetic field by small-scale turbulent disturbances.
Finally, the $\gamma$ term is a different kind of magnetic field amplification mechanism driven by the cross-helicity effect.
It was first proposed by \cite{Yoshizawa:1990} and is also used by \cite{Sur+Brandenburg:2009,Yokoi:2013}.
This effect operates if cross-helicity is present in the magnetohydrodynamic description of turbulence.
The cross-helicity describes the linkeage between magnetic field and vortex flows.
Like the $\beta$ term, the $\gamma$ also contains a curl of a background field, here of the flow velocity: $\vec{\Omega}_0 = \vec{\nabla} \times \vec{U}_0$.

%%%%%%%%%%%%%%%%%%%%%
\subsection{Transport coefficients}

Various models of the transport coefficients $\alpha$ and $\beta$ are possible.
An early formulation of $\alpha$ and $\beta$ was derived from the mean-field electrodynamics by \cite{Krause+Rädler:1980}:
\begin{eqnarray}
\alpha &=& \textstyle{\frac{1}{3}} \tau \langle -\delta \vec{U} \cdot \delta \vec{\Omega} \rangle \nonumber
\\
\beta &=& \textstyle{\frac{1}{3}} \tau \langle \delta \vec{U} \cdot \delta \vec{U} \rangle \label{eq:alpha1beta1}
\end{eqnarray}
$\tau$ denotes the characteristic time scale of the fluctuations and describes the half-time of the decay of the turbulent energy and the helical strucures.
Analogous to $\alpha$ and $\beta$ we add the $\gamma$ transport coefficient to our model as follows:
\begin{equation}
\gamma = \textstyle{\frac{1}{3}} \tau \langle \delta \vec{U} \cdot \delta \vec{B} \rangle
\label{eq:gamma1}
\end{equation}

A second form gives a more general modeling for the transport coefficients; see Eq.\,(23)--(25) in \cite{Yoshizawa:1990} and Eq.\,36(a)--(c) in \cite{Yokoi:2013}.
In particular, for turbulent flows with high magnetic Reynolds or Lundquist numbers, a dependency on the current helicity appears in the $\alpha$ coefficient \citep{Pouquet+al:1976}:
\begin{eqnarray}
\alpha &=& C_\alpha \tau ~ \langle - \delta \vec{u} \cdot \delta \vec{\Omega} + \delta \vec{b} \cdot \delta \vec{J} \rangle \label{eq:alpha2}
\\
\beta  &=& C_\beta  \tau ~ \textstyle{\frac{1}{2}} \langle {\vert \delta \vec{u} \vert}^2 + {\vert \delta \vec{b}\vert}^2 \rangle \label{eq:beta2}
\\
\gamma &=& C_\gamma \tau ~ \langle \delta \vec{u} \cdot \delta \vec{b} \rangle \label{eq:gamma2}
\end{eqnarray}
The magnetic field is normalized here to Alfv{\'e}n velocity units: $\vec{b} = \vec{B} / \sqrt{\mu_0 \rho_0}$ and the plasma flow velocity as $\vec{u} = \vec{U}$.
The coefficients $C_\alpha$, $C_\beta$, and $C_\gamma$ are model parameters.
They were estimated by \cite{Hamba:1992} as $C_\alpha = O(10^{-2})$, $C_\beta = O(10^{-1})$, and $C_\gamma = O(10^{-1})$ for the turbulent electromotive force.

The structure of this work is as following: in Sect.\,\ref{methods} we use Taylor's frozen-in flow hypothesis \citep{Taylor:1938} to reduce the curl to one-dimensional derivatives and calculate the transport coefficients from Eq.\,(\ref{eq:alpha1beta1}).
Then we construct a shock-front model of a helical magnetic field and a vortex in the plasma flow velocity, which corresponds to a field configuration as expected from kinetic helicity.
With an initial estimate of free model parameters we find specific signatures in the $\alpha$ and $\beta$ terms.
In Sect.\,\ref{inner} we employ our method to the spacecraft data and compare the resulting signatures to the model prediction.

%%%%%%%%%%%%%%%%%%%%%%%%%%%%%%%%%%%%%%%%%%%%
\section{Algorithm and model construction} \label{methods}

%%%%%%%%%%%%%%%%%%%%%
\subsection{Taylor's hypothesis} \label{taylor}

The difficulty in evaluating the transport coefficients lies in the task of determining the spatial derivatives from spacecraft data in order to calculate the vorticity $\vec{\Omega} = \vec{\nabla} \times \vec{U}$ in the plasma bulk flow, where $\vec{U}$ is a function of space and time.
Also the electric current density is defined by spatial derivatives of the magnetic field data through Amp{\`e}re's law: $\mu_0\vec{J} = \vec{\nabla} \times \vec{B}$.
The spatial derivatives can be obtained from multi-spacecraft mission data using four spacecraft in a tetrahedral formation, like {\em Cluster} \citep{Balogh+al:1997} or {\em MMS} \citep{Torbert+al:2016}.
Such data allow to separate time derivatives from spatial derivatives within the three-dimensional (3D) vectors of the magnetic field and plasma flow velocity.
Unfortunately, the above-mentioned spacecraft orbit around the Earth and do not provide data from the heliosphere below Earth orbit.
Therefore, we use data from the {\em Helios} spacecraft that provide information on the proton velocity, density, and the magnetic field down to a distance of $0.3\,\rm{au}$ from the Sun.
However, the two {\em Helios} spacecraft do not orbit in formation, so we can only use single-spacecraft data and the field vectors depend only on time and have no spatial derivative.
To overcome this issue, we use a streamwise coordinate system, with the coordinate $z$ in the direction to the mean flow.

We obtain a spatial derivative from the time series using Taylor's frozen-in flow hypothesis \citep{Taylor:1938}.
If we assume that $\frac{\delta U}{U_0} \ll 1$, the turbulent eddies evolve on a time scale larger than the observed event and thus the proper change in their structure is negligible.
We can then replace the spatial derivative along $z$ with the time derivative: $\partial z = U_0 \partial t$.
This leads to a reduced form of the nabla operator that only consists of the $z$ derivative:
\begin{equation}
\vec{\nabla} = \left( \begin{array}{c} \partial / \partial x \\ \partial / \partial y \\ \partial / \partial z \end{array} \right) \to \frac{1}{U_0} \left( \begin{array}{c} 0 \\ 0 \\ \partial / \partial t \end{array} \right).
\label{eq:curl}
\end{equation}
The curl of a function $h$ then becomes:
\begin{equation}
\vec{\nabla} \times \left( \begin{array}{c} h_x \\ h_y \\ h_z \end{array} \right) \to \frac{1}{U_0} \left( \begin{array}{c} - \partial h_y / \partial t \\ + \partial h_x / \partial t \\ 0 \end{array} \right).
\label{eq:curl2}
\end{equation}

%%%%%%%%%%%%%%%%%%%%%
\subsection{Separation of background and fluctuation} \label{background}

In large-eddy simulations (LESs) any quantity can be separated into a resolved grid scale (GS) and an unresolved sub-grid scale (SGS) part.
This is possible, e.g., with a spatial averaging, a time averaging, or with a frequency-domain filtering method.
The GS then represents the homogenous mean field while the SGS contains the inhomogenous turbulent part.
For in-situ measurements of the solar wind, the SGS quantity cannot be obtained for a single event because the fluctuations are of spatial scales larger than the resolved GS.
If we applied here the same method as used for LESs, the curl of our mean fields would either vanish or become negligible -- and consequently only the $\alpha$ term would give significant non-zero contributions.
Therefore, we require an alternative formulation of the $\beta$ and $\gamma$ terms that contain such a curl.

For our CME model, we consider the Parker spiral as the background field, which is relatively constant while the magnetic transient event is the fluctuation that passes by the spacecraft.
From the observational data, we determine the mean field through a moving Gaussian convolution over large spatial scales and obtain the small-scale fluctuations as the residual after subtracting the obtained mean field.
The Gaussian-convolution filtered signal, as well as its residual, contains contributions from scales larger and smaller than the characteristic length scale used for the filtering, which is set via $\sigma_0$ in Sect.\,\ref{results2}.
By this method, the obtained fluctuations $\delta \vec{B}$ and $\delta \vec{U}$ both still contain parts of the background (or mean) fields.

\begin{figure}
\begin{center}
\includegraphics[trim=0 0 0 0,width=8.0cm]{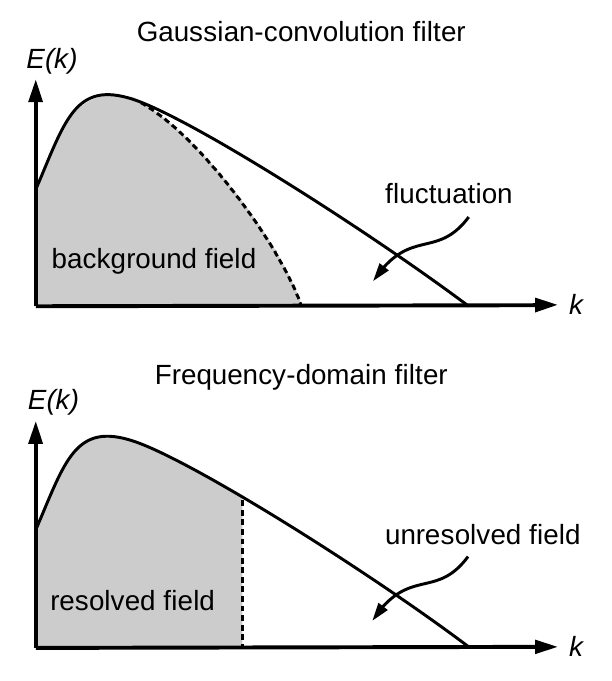}
\end{center}
\caption{Spectrum of a Gaussian-convolution filtered signal (upper panel) and a frequency-based filtering (lower panel).
The large-scale background or resolved field (gray shaded), as well as the residual fluctuation or unresolved field, have different contributions from different scales.}
\label{fig:scale}
\end{figure}

In contrast to the Gaussian-convolution separation described above, a frequency-based filtering method would split those contributions strictly at the characteristic scale, as often applied in turbulence research and LESs, see Fig.\,\ref{fig:scale}.
This leads to different fluctuations $\delta \vec{B}$ and $\delta \vec{U}$ than obtained through the Gaussian convolution.
Also the exact choice of this characteristic separation scale has an influence on the result.
It is worth investigating, in future work, how the separation method for background field and fluctuation could be improved for in-situ observations.
One possibility may be to split the Gaussian-convolution residuals further with a frequency-based filter into their resolved GS and unresolved SGS components.
This would allow us to formulate the electromotive force more consistently with previous works in turbulent-transport theory.

%%%%%%%%%%%%%%%%%%%%%
\subsection{Alternative formulation of electromotive force} \label{em-force}

In the following, instead of the curl of the background quantities $\vec{\nabla} \times \vec{B}_0$ and $\vec{\nabla} \times \vec{U}_0$ that would both vanish here, we now use the curl of the Gaussian-convolution residuals; see Sect.\,\ref{background}.
The contributions from larger and smaller scales contained in these residuals still have sufficiently large derivatives to get a non-zero current density and vorticity with the curls of $\delta \vec{B}$ and $\delta \vec{U}$.
This leads us from Eq.\,\ref{eq:em2} to an alternative formulation of the electromotive force $\vec{M}_2$ as:
\begin{equation}
\vec{M}_2 = \alpha \vec{B}_0 - \beta (\vec{\nabla} \times \delta \vec{B}) + \gamma (\vec{\nabla} \times \delta \vec{U})
\label{eq:em_2_raw}
\end{equation}

We can now insert the transport coefficients from Eqs.\,(\ref{eq:alpha1beta1}) and (\ref{eq:gamma1}) into (\ref{eq:em_2_raw}) and get:
\begin{eqnarray}
\vec{M}_2 = &~& \textstyle{\frac{1}{3}} \tau \langle -\delta \vec{U} \cdot (\vec{\nabla} \times \delta \vec{U}) \rangle \vec{B}_0 \nonumber \\
&-& \textstyle{\frac{1}{3}} \tau \langle \delta \vec{U} \cdot \delta \vec{U} \rangle (\vec{\nabla} \times \delta \vec{B}) \nonumber \\
&+& \textstyle{\frac{1}{3}} \tau \langle \delta \vec{U} \cdot \delta \vec{B} \rangle (\vec{\nabla} \times \delta \vec{U}) \label{eq:em_2}
\end{eqnarray}
where $\tau$ is the time scale used for the averaging in order to determine the observed background fields, see Sect.\,\ref{results2}.

%%%%%%%%%%%%%%%%%%%%%
\subsection{Model construction} \label{model}

Realistic turbulence models usually need to contain the production and dissipation mechanisms on large and small scales.
However, in turbulent systems the large scales contain the energy, which determines the turbulent transport properties.
In the region of the heliosphere that we analyze, from 0.4\,\rm{au}, the solar wind is no longer accelerated and starts to decay.
We expect that the injection of energy on the largest scales starts to fade out and that the dissipation begins to dominate.
This allows us to construct a simplified model, where the turbulent transport coefficients may be described also while neglecting the production mechanism.

We use a streamwise coordinate system with the third component aligned with the mean flow $\vec{U}_0$.
Due to the proximity to the Sun (about 0.4\,\rm{au}) the Parker spiral is still rather radial.
Therefore, we assume the mean magnetic field $\vec{B}_0$ to be parallel with $\vec{U}_0$.
Alltogether, we define the large-scale solar wind background, representing a quiet solar wind, as
\begin{equation}
\vec{B}_0 = \left( \begin{array}{c} 0 \\ 0 \\ B_{0} \end{array} \right),
\vec{U}_0 = \left( \begin{array}{c} 0 \\ 0 \\ U_{0} \end{array} \right).
\end{equation}

We define a second set of fields that represent the perturbation caused by a magnetic transient event.
This region contains a helically twisted magnetic field $\vec{B}_1$ and a vortex in the plasma flow velocity $\vec{U}_1$.
The construction is made so that the large-scale fields ($\vec{B}_0$, $\vec{U}_0$) and small-scale fields ($\vec{B}_1$, $\vec{U}_1$) are of the same order of magnitude.

The model contains several free parameters:
$B_0$ and $U_0$ are the strength of the mean magnetic field and mean plasma flow velocity.

From these background values, we expect the mean fields to increase to a plateau value before and after a magnetic transient event, which we express through $\epsilon_x, \epsilon_y, \epsilon_z$ and $\iota_x, \iota_y, \iota_z$.
Additionally, we define a peak value that sets the amplitude of the fields for the center of the vortex structure through $\psi_x, \psi_y, \psi_z$ and $\phi_x, \phi_y, \phi_z$.
The direction of the field vectors depends on $\Theta(y,z)=\textrm{atan(y/z)}$.
$g_B$ describes the shape of the peak in the magnetic field, while we use ${f_p}$ for the non-diverging center of the velocity vortex, see Eqs.\,\ref{eq:profile} and \ref{eq:gauss}.
We define the disturbed solar-wind field vectors as:
\begin{eqnarray}
\vec{B}_1 &=& B_{0} \left [ \left ( \begin{array}{@{}r@{}l@{}} {\epsilon}_{x}& \\ - {\epsilon}_{y}& \sin{\Theta} \\ {\epsilon}_{z}& \cos{\Theta} \end{array} \right ) + \left ( \begin{array}{@{}r@{}l@{}} {\psi}_{x}& \\ - {\psi}_{y}& \sin{\Theta} \\ {\psi}_{z}& \cos{\Theta} \end{array} \right ) g_B \right ] \nonumber \\ \label{eq:helical_b1}
\\
\vec{U}_1 &=& U_{0} \left [ \left( \begin{array}{c} {\iota}_{x} \\ {\iota}_{y} \\ {\iota}_{z} \end{array} \right) + \left ( \begin{array}{@{}r@{}l@{}} {\phi}_{x}& \\ {\phi}_{y}& (\cos{\Theta} + \sin{\Theta}) \\ {\phi}_{z}& (\cos{\Theta} - \sin{\Theta}) \end{array} \right ) \vert f_p \vert \right ] \nonumber \\ \label{eq:helical_u1}
\end{eqnarray}

We now focus on a one-dimensional cut through the center of the vortex.
This means, when a 3D magnetic shock front passes by a spacecraft, one always observes the innermost structure of that shock front.
Therefore, we may choose $\cos{\Theta} = 0$ for all times $t$ and replace $\sin{\Theta}$ with
\begin{equation}
\sin{\Theta} = \left \{
\begin{array}{rcl}
-1 &:& t < 0 \\
 1 &:& t > 0 \\
 0 &:& t = 0
\end{array}
\right .
\label{eq:theta}
\end{equation}
where $t=0$ specifies the center of the shock front in the time series.
However, to avoid that $\vec{B}_1$ contains a non-steady step function with diverging derivatives at the vortex center, we use a smooth transition instead, here a fifth-order polynomial with zero first and second derivatives at the boundaries:
\begin{equation}
s_t \left( \xi = \frac{t}{c_g} \right) = \frac{1}{2} + \frac{15}{16} \xi - \frac{5}{8} \xi^3 + \frac{3}{16} \xi^5
\label{eq:sign}
\end{equation}
with $\vert t \vert \le c_g$ and $c_g$ as the width of the transition.
The coefficients of the terms in $s_t$ follow from the boundary conditions $s_t(-1)=0$, $s_t(0)=1/2$, and $s_t(1)=1$.

With the assumption of incompressibility the radial profile of a vortex asympthotically decreases as $1/R$, with $R(t)=\vert t \vert$.
This would lead to a diverging velocity at the vortex center $R \to 0$.
But, for a realistic plasma vortex motion, the velocity at the center should vanish.
Therefore, we replace $1/R$ with the radial profile
\begin{equation}
f_p(t,c_f) = \frac{t}{\textrm{cosh}(t/c_f)},
\label{eq:profile}
\end{equation}
where $c_f$ is another free parameter that limits the derivatives at the vortex center.

We amplify the helical magnetic field with a Gaussian profile to mimic a peak in the magnetic field around the shock front:
\begin{equation}
g_B(t,\sigma_B) = \frac{1}{\sqrt{2 \pi {\sigma_B}^2}} \exp{\left( -\frac{t^2}{2{\sigma_B}^2} \right)},
\label{eq:gauss}
\end{equation}
where $\sigma_B$ sets the width of the peak and completes our set of free parameters.

The fluctuating fields can now be written as:
\begin{eqnarray}
\vec{B}_1 &=& B_{0} \left ( \begin{array}{@{}r@{}l@{}} {\epsilon}_{x} +& {\psi}_{x} g_B \\ - ( {\epsilon}_{y} +& {\psi}_{y} g_B) \left(2 s_t(\frac{t}{c_g}) - 1\right) \\  &0  \end{array} \right ) \label{eq:helical_b}
\\
\vec{U}_1 &=& U_{0} \left ( \begin{array}{@{}r@{}l@{}} {\iota}_{x} +& {\phi}_{x} \vert f_p \vert \\ {\iota}_{y} +& {\phi}_{y} f_p \\  {\iota}_{z} -& {\phi}_{z} f_p \end{array} \right ) \label{eq:helical_u}
\end{eqnarray}

\begin{figure}
\begin{center}
\includegraphics[trim=0 0 0 0,width=8.0cm]{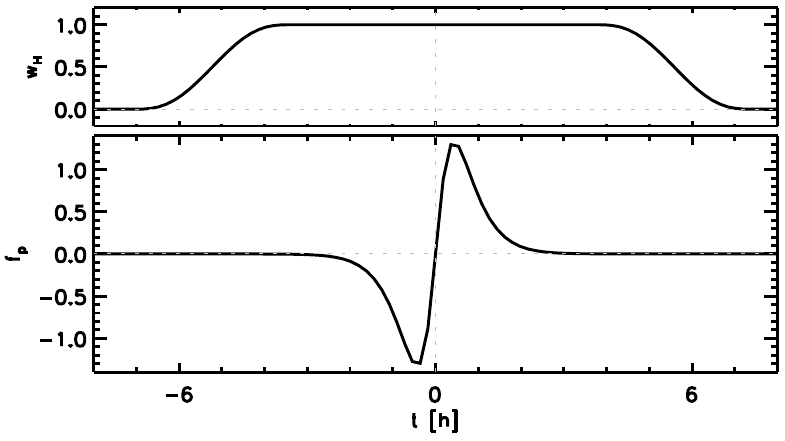}
\end{center}
\caption{Upper panel: weight function for the helical field phase $w_H$ ($c_h=1.8$\,hr).
Lower panel: radial profile of the vortex $f_p$ ($c_f=0.36$\,hr).
The horizontal axis is the time distance to the vortex center in hours.}
\label{fig:prof_weight}
\end{figure}

As a final step, we use a weight function $w_H$ to define a specific interval during which the helical fields dominate.
This allows us to switch smoothly between quiet and helical solar wind with the smooth transition $s_t$ defined in Eq.\,(\ref{eq:sign}):
\begin{equation}
w_H = \left \{
\begin{array}{rcrl}
    0                &:&               t &< -\Delta-c_h \\
  s_t(\frac{t}{c_h}) &:& -\Delta-c_h < t &< -\Delta \\
    1                &:&     -\Delta < t &< \Delta \\
1-s_t(\frac{t}{c_h}) &:&      \Delta < t &< \Delta+c_h \\
    0                &:&  \Delta+c_h < t &
\end{array}
\right .
\label{eq:weight}
\end{equation}
with $c_h = 1.8$\,hr and $\Delta= 3.6$\,hr.
In Fig.\,\ref{fig:prof_weight} we show $w_H$ and the radial profile $f_p$.

The total vector fields are now given by the sums of the smoothly tapered background and helical fields:
\begin{eqnarray}
\vec{B}_{tot} =& (1 - \kappa_B w_H) \vec{B}_0 + w_H \vec{B}_1 \label{eq:total_b}
\\
\vec{U}_{tot} =& (1 - \kappa_U w_H) \vec{U}_0 + w_H \vec{U}_1 \label{eq:total_u}
\end{eqnarray}

For this study we choose $\kappa_B = \kappa_U = 0$, meaning that the background fields remain constantly present throughout the whole event, while $\vec{B}_1$ and $\vec{U}_1$ describe the distrubances.
This allows us to take $\vec{B}_0$ and $\vec{U}_0$ as the mean fields, as well as $\vec{B}_1$ and $\vec{U}_1$ as the fluctuations.
The transport coefficients and the electromotive force can then be expressed with our model parameters as:
\begin{equation}
\alpha = -\frac{\tau}{3} U_0 \iota_x \phi_y {w_H}^2 \frac{\partial f_p}{\partial t}
\label{eq:sol_alpha}
\end{equation}
\begin{eqnarray}
\beta =& \frac{\tau}{3} {U_0}^2 {w_H}^2 \left ( {\iota_x}^2 + {\iota_y}^2 + {\iota_z}^2 + 2 \iota_y \phi_y f_p \right. \nonumber \\ & \left. - 2 \iota_z \phi_z f_p + {\phi_y}^2 {f_p}^2 + {\phi_z}^2 {f_p}^2 \right )
\label{eq:sol_beta}
\end{eqnarray}
\begin{eqnarray}
\gamma =& \frac{\tau}{3} U_0 B_0 {w_H}^2 \left [ \iota_x ( \epsilon_x + \psi_x g_B) \right. \nonumber \\ & \left. - ( \iota_y + \phi_y f_p ) ( \epsilon_y s_t + \psi_y s_t g_B ) \right ]
\label{eq:sol_gamma}
\end{eqnarray}

\begin{eqnarray}
M_1 &=& U_0 B_0 {w_H}^2 \left \{ \left ( \iota_z - \phi_z f_p \right )^2 \left ( \epsilon_y s_t + \psi_y s_t g_B \right )^2 \right. \nonumber \\
&& \left. + \left [ \iota_x ( \epsilon_y s_t + \psi_y s_t g_B ) \right. \right. \nonumber \\
&& ~~~~ \left. \left. + ( \iota_y + \phi_y f_p ) ( \epsilon_x + \psi_x g_B ) \right ]^2 \right. \nonumber \\
&& \left. + \left ( \iota_z - \phi_z f_p )^2 ( \epsilon_x + \psi_x g_B \right )^2 \right \}^{\frac{1}{2}}
\label{eq:sol_em}
\end{eqnarray}
\begin{eqnarray}
M_2 &=& \left \{ \left. (\alpha B_0)^2 \right. \right. \nonumber \\
&& \left. + \left [ - \beta \frac{B_0}{U_0} \left ( \epsilon_y \frac{\partial w_H s_t}{\partial t} + \psi_y \frac{\partial w_H s_t g_B}{\partial t} \right ) \right. \right. \nonumber \\
&& ~~~~ \left. \left. + \gamma \left ( \iota_y \frac{\partial w_H}{\partial t} + \phi_y \frac{\partial w_H f_p}{\partial t} \right ) \right ]^2 \right. \nonumber \\
&& \left. + \left [ - \beta \frac{B_0}{U_0} \left ( \epsilon_x \frac{\partial w_H}{\partial t} + \psi_x \frac{\partial w_H g_B}{\partial t} \right ) \right. \right. \nonumber \\
&& ~~~~ \left. \left. + \gamma \left ( \iota_x \frac{\partial w_H}{\partial t} \right ) \right ]^2 \right \}^{\frac{1}{2}}
\label{eq:sol_em2}
\end{eqnarray}

%%%%%%%%%%%%%%%%%%%%%
\subsection{Model parameters} \label{results}

\begin{figure}
\begin{center}
\includegraphics[trim=0 0 0 0,width=8.0cm]{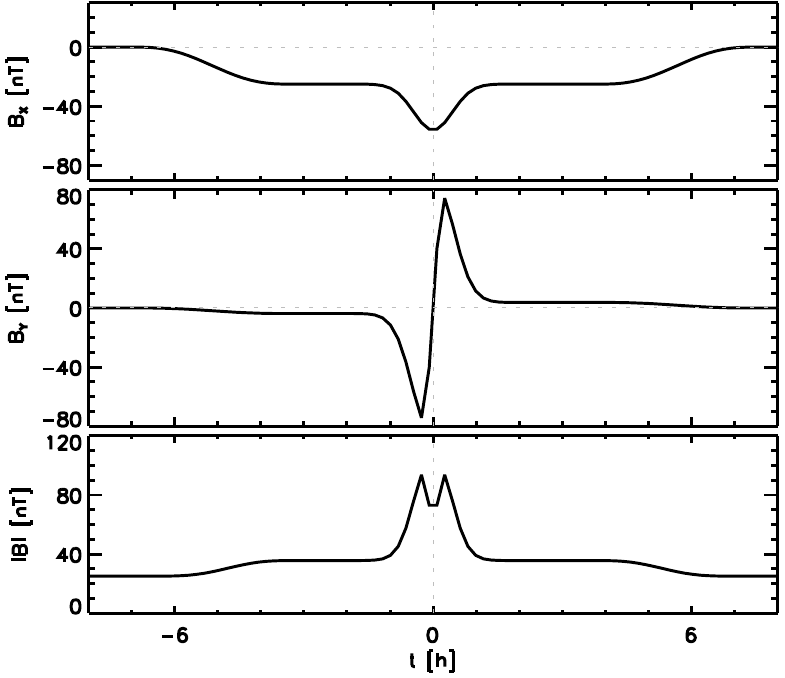}
\end{center}
\caption{Helical magnetic field components and magnitude of the magnetic field from the vortex model.
The radial component is not shown as it is constant at $B_0=25\,\rm{nT}$.
The x-axis shows the time distance to the vortex center in hours.}
\label{fig:b_model1}
\end{figure}
\begin{figure}
\begin{center}
\includegraphics[trim=0 0 0 0,width=8.0cm]{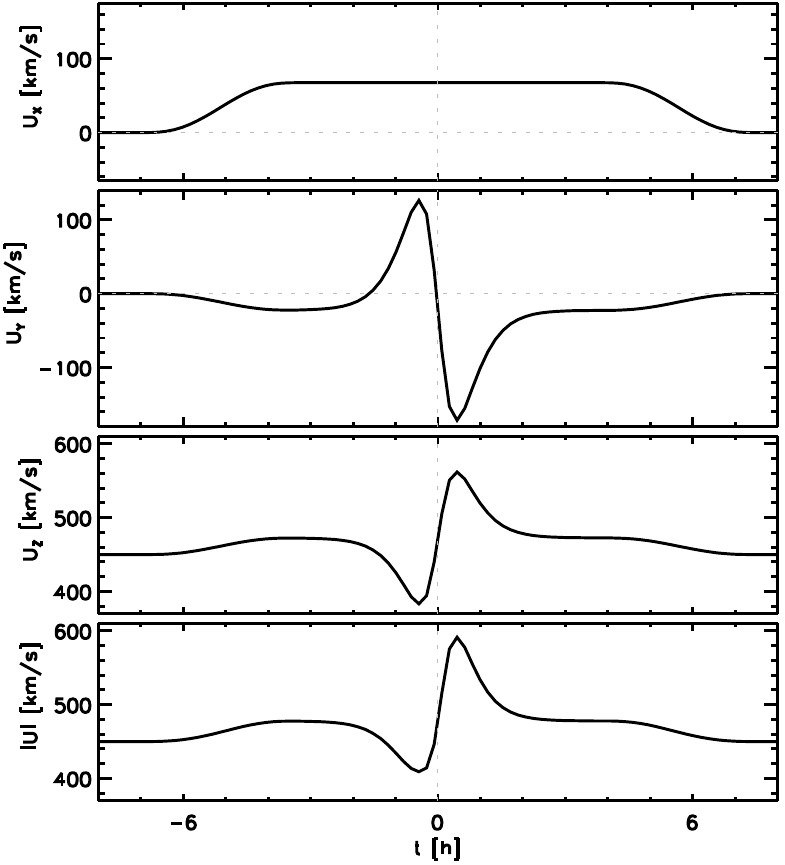}
\end{center}
\caption{Plasma flow velocity components and magnitude of the plasma flow velocity from the vortex model.
The x-axis shows the time distance to the vortex center in hours.}
\label{fig:u_model1}
\end{figure}

As the $z$ component of the helical magnetic field is always zero, see Eq.\,(\ref{eq:helical_b}), the two parameters ${\epsilon}_z$ and ${\psi}_z$ can be removed from our model.
Without loss of generality we assume a vortex in the $y$--$z$ plane and hence set ${\phi}_x=0$.
The other free parameters are chosen by an educated guess: we estimate the strength of the large-scale fields in the solar wind as $B_{\rm{SW}} = 25\,\rm{nT}$ and $U_{\rm{SW}} = 450\,\rm{km/s}$.
With ${\epsilon}_x = -1$, we assume a helical magnetic field of the same magnitude as the solar wind background field in the $x$-direction and a smaller amplitude in the $y$-direction, ${\epsilon}_y = -0.15$.
We set the peak values of the helical magnetic field by the parameters ${\psi}_x = -1.25$ and ${\psi}_y = -3.5$.
For the plasma flow velocity, we assume that there is a small component normal to the vortex ${\iota}_x = 0.15$ and a three times smaller component along $y$ and $z$: ${\iota}_y=-0,05$, ${\iota}_z = 0.05$.
We choose the amplitude of the plasma flow vortex as ${\phi}_y = -0.25$ and ${\phi}_z = -0.15$.
The broadness parameters of $g_B$, $f_p$, $s_t$, and $w_H$ we estimate to be $\sigma_B=0.45$\,hr, $c_f=0.36$\,hr, $c_g=0.36$\,hr, and $c_h=1.8$\,hr and the center of the vortex ($t=0$) we set at 1978 April 18, 18:25:38 UT.
The time resolution $\Delta t=648$\,s is chosen to be identical to the time resolution of the {\em Helios} data (Sect.\,\ref{helios}).
The resulting magnetic and velocity fields are plotted in Figs.\,\ref{fig:b_model1} and\,\ref{fig:u_model1}.

Our initial choice for the characteristic time scale of the fluctuations is $\tau = 7.5 \Delta t = 1.35\,\rm{hr}$.
This time scale is chosen with respect to the evaluation of the background fields from the {\em Helios} spacecraft data; see Sect.\,\ref{results2}.

We can now compute the transport coefficients from Eqs.\,(\ref{eq:sol_alpha})--(\ref{eq:sol_gamma}).
In the upper panel of Fig.\,\ref{fig:alpha_beta_1} we find a single peak in the kinetic helicity ($\alpha$) surrounded by two smaller peaks with opposite sign.
However, the smaller peaks come from the derivative of the radial profile and do not necessarily mean that there is instantaneous dynamo action in the opposite sense.

The turbulent diffusion ($\beta$) features a double peak in the center of the vortex with the left peak being smaller than the right peak; see the lower panel of Fig.\,\ref{fig:alpha_beta_1}.
The double peak in $\beta$ is formed by the radial profile of the vortex $f_p$; the different amplitudes of the peaks are caused by the ratio between the plateaus and peak values in the plasma flow velocity ($\iota_y$, $\iota_z$, $\phi_y$, $\phi_z$).
Unlike the $\alpha$ term, the turbulent diffusion is not strongest at the vortex center, because the magnetic field fluctuation disappears there.
The plateau in $\beta$ is mainly formed by the $x$ component of the plasma flow normal to the vortex plane.
This normal component needs to be different from zero, so that the $\alpha$ transport coefficient may exist in our model.

The cross-helicity ($\gamma$) term resembles the $\beta$ term, but with an opposite sign.
There is a plateau and a double peak at the vortex center with the left peak being smaller than the right peak.
However, the asymmetry between both peaks in the $\gamma$ term is less strong than for the $\beta$ transport coefficient of the model.
When the sign is opposite to the $\beta$ term, this reflects the magnetic polarity and, in particular, that the magnetic field vector is antiparallel to the plasma flow velocity.
If both vectors are parallel, one expects the same sign for $\beta$ and $\gamma$.

The magnitude of the electromotive force $M_1$ (Fig.\,\ref{fig:em_comp_obs}) calculated from mean-field electrodynamics (Eq.\,(\ref{eq:sol_em})) features a plateau and a double peak around the vortex center similar to the turbulent diffusion term with the left peak being smaller than the right peak.
Here, the sign of the plateau parameters in the plasma flow velocity ($\iota_y$, $\iota_z$) determines which peak is higher.

On the other hand, if we use Eq.\,(\ref{eq:sol_em2}) to calculate the magnitude of the electromotive force from the turbulent transport coefficients ($M_2$), we find a superposition of the three peaks with an enhanced amplitude on the right side.
The plateaus in the $\beta$ and $\gamma$ coefficients do not carry over to the model electromotive force $M_2$ because there are no plateaus in the current density and vorticity of our model.
This is similar to the observational data, where the plateau phase in $M_2$ is less significant than in $M_1$; see the logarithmic plot in Fig.\,\ref{fig:em_comp_obs_log}.

When we compare both formulations of the electromotive force ($M_1$ and $M_2$), the double peaks visible in $M_1$ seem to be merged in $M_2$; see the red dashed line in the lower panel of Fig.\,\ref{fig:em_comp_obs}.
The reason for this less significant double-peak structure in $M_2$ is not solely that the $\alpha$ coefficient shows a single peak in the vortex center.
Indeed, it is the contribution to $M_2$ of the whole $\gamma$ term that features a strong single peak at the vortex center, which overlays the still significant double-peak structure obtained from the sum of the $\alpha$ and $\beta$ terms.
The $\gamma$ term itself is dominated by the vorticity that has a strong single peak in the vortex center due to the large derivatives in the velocity field.

%%%%%%%%%%%%%%%%%%%%%%%%%%%%%%%%%%%%%%%%%%%%
\section{Application in the inner heliosphere} \label{inner}

%%%%%%%%%%%%%%%%%%%%%
\subsection{{\em Helios} data overview} \label{helios}

We use magnetic field and proton velocities obtained by the {\em Helios-2} spacecraft \citep{Schwenn+al:1975,Musmann+al:1975} from 1978 April 18--19.
At that point in time, {\em Helios-2} was located at a distance of 0.41\,\rm{au} from the Sun.
The {\em Helios} data are in the Spacecraft Solar Ecliptic (SSE) coordinate system: the $XY$-plane is the Earth mean ecliptic of date and the $+X$-axis is the projection of the vector spacecraft--Sun on the $XY$-plane.
The $+Z$-axis is the direction of the ecliptic south pole \citep{Fränz+Harper:2002}.
According to the header of the data files, the magnetic field and plasma flow velocity are given in radial--tangential--normal (RTN) coordinates with $R=-X$, $T=-Y$, and $N=+Z$.
There was no correction made regarding the RTN system being defined relative to the helioequatorial plane instead of the ecliptic plane, which results in errors up to $1\%$.
The magnetic field and plasma flow velocity, however, are consistent with each other and can be compared directly; see the Acknowlegements for the link to the {\em Helios} data.
These data have a sampling rate of $40.5\,\rm{s}$, which is the highest rate that is accessible for both the magnetic field and the proton flow velocity.
The magnetic field data are obtainable at a higher sampling rate, but here they are reduced to the same resolution as the proton flow velocity.
However, the sampling rate is not constant over the entire interval, because of changes between different observing modes, plus there are data gaps.
These irregularities in the data we reduce by binning the data to an equidistant time discretization.
As we observe a time interval with a magnetic cloud passing by, we choose the sampling rate at the shock front ($\Delta t = 648\,\rm{s}$) as the resolution for the whole interval.
We replace single missing data points by a quadratic interpolation.

%%%%%%%%%%%%%%%%%%%%%
\subsection{Observational results} \label{results2}

We perform a coordinate transformation to streamwise coordinates using the median values of the three components of the plasma flow velocity.
This turns our coordinate system into the direction of the mean flow.
In this case, however, as the radial component before the transformation is in the direction of the Sun, the resulting coordinate system is very close to the initial one, therefore we persist with the RTN notation.
To determine the mean magnetic field and the mean plasma flow velocity, we use a smoothing function with a Gaussian kernel with a standard deviation $\sigma_0$ of 11 data points.
This $\sigma_0$ mimics a box-car smoothing with a moving average over 30 neighboring points, but avoids the unwanted artifacts of this smoothing, such as those due to outliers in the data.
The fluctuation we calculate as the difference of the original and the mean field.
For the half-time of the fluctuations $\tau$, we assume that $4 \tau$ is the minimum time for the decay of turbulent structures to a negligible value, here $1/2^4$.
With that, we obtain a characteristic time scale of one fourth of the smoothing value: $\tau = 30/4 \Delta t = 1.35\,\rm{hr}$.

\begin{figure}
\begin{center}
\includegraphics[trim=0 0 0 0,width=8.0cm]{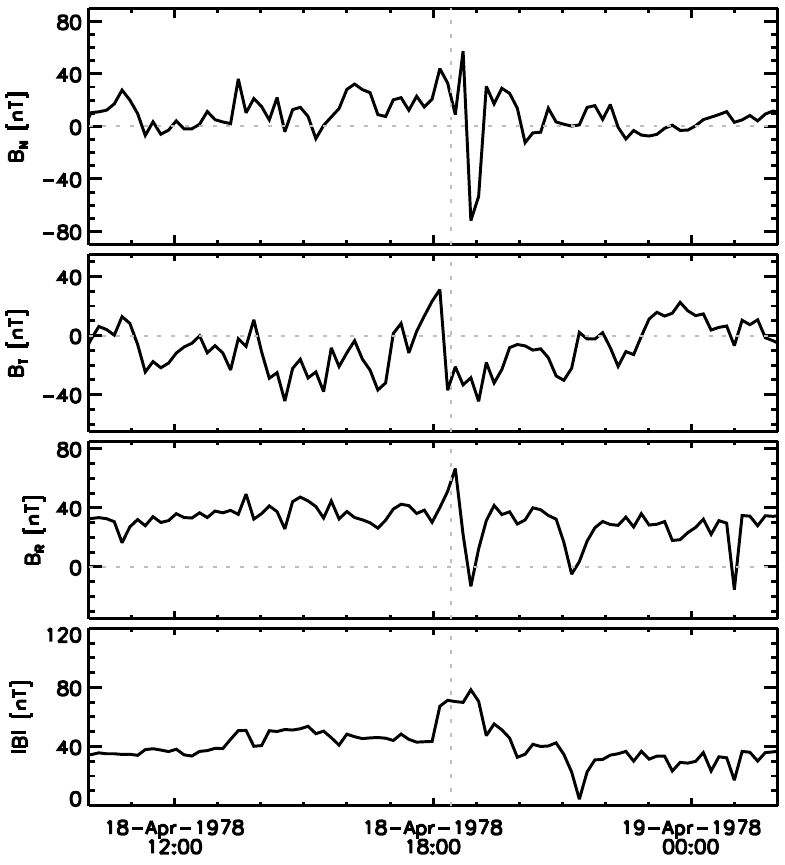}
\end{center}
\caption{{\em Helios-2} observation of the magnetic field with derived components ($R,T,N$).
The timestamp is in UT.}
\label{fig:b_obs}
\end{figure}
\begin{figure}
\begin{center}
\includegraphics[trim=0 0 0 0,width=8.0cm]{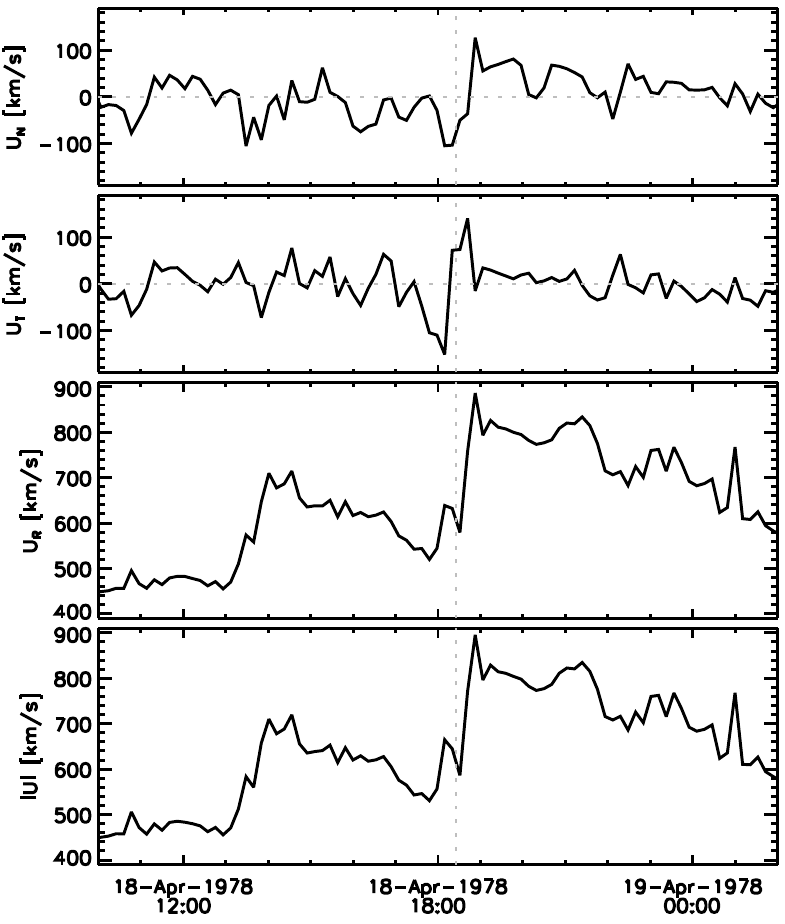}
\end{center}
\caption{{\em Helios-2} observation of the proton velocity with derived components ($R,T,N$).
The timestamp is in UT.}
\label{fig:u_obs}
\end{figure}

\begin{figure}
\begin{center}
\includegraphics[trim=0 0 0 0,width=8.0cm]{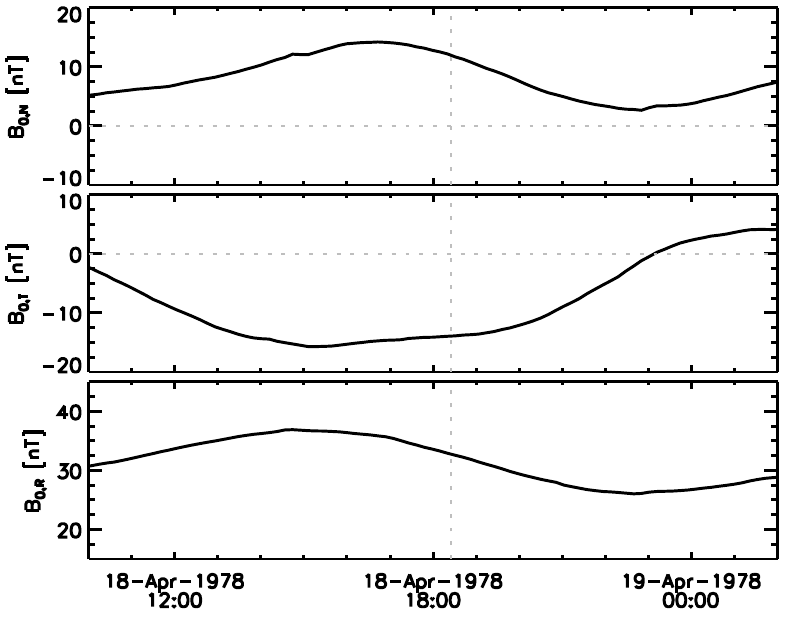}
\end{center}
\caption{Mean magnetic field components as derived from the {\em Helios-2} observations.
The timestamp is in UT.}
\label{fig:b_obs_m}
\end{figure}
\begin{figure}
\begin{center}
\includegraphics[trim=0 0 0 0,width=8.0cm]{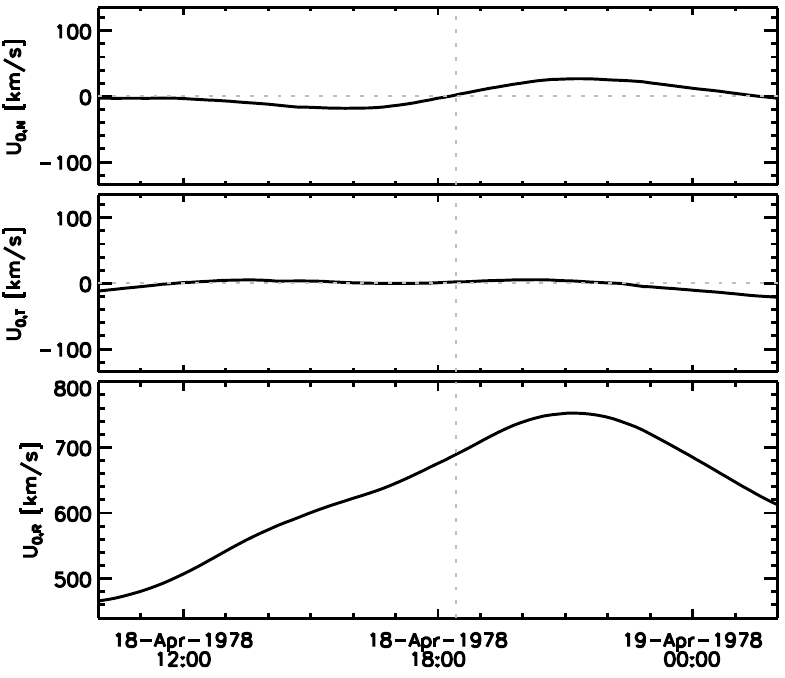}
\end{center}
\caption{Mean proton flow velocity components as derived from the {\em Helios-2} observations.
The timestamp is in UT.}
\label{fig:u_obs_m}
\end{figure}

\begin{figure}
\begin{center}
\includegraphics[trim=0 0 0 0,width=8.0cm]{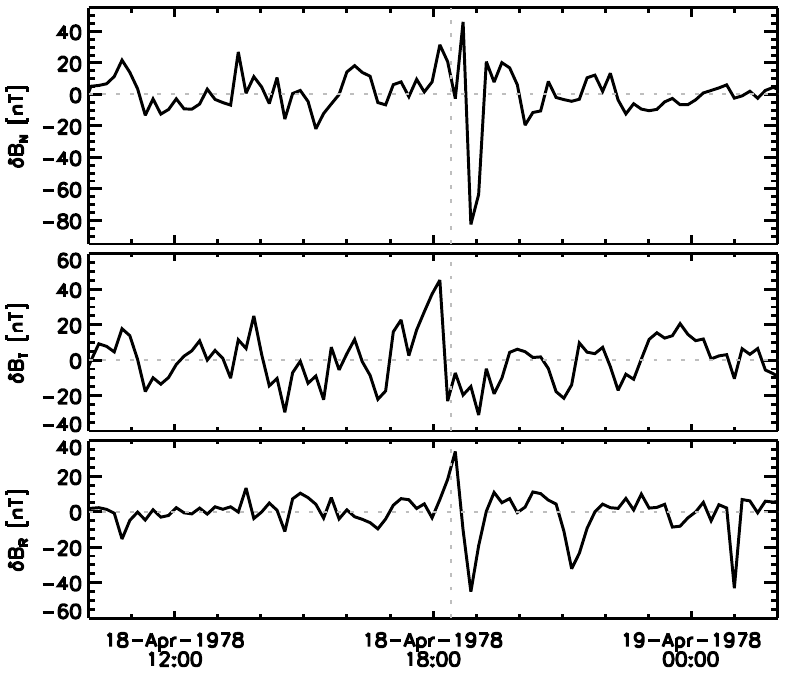}
\end{center}
\caption{Fluctuating magnetic field components as derived from the {\em Helios-2}  observations.
The timestamp is in UT.}
\label{fig:b_obs_f}
\end{figure}

\begin{figure}
\begin{center}
\includegraphics[trim=0 0 0 0,width=8.0cm]{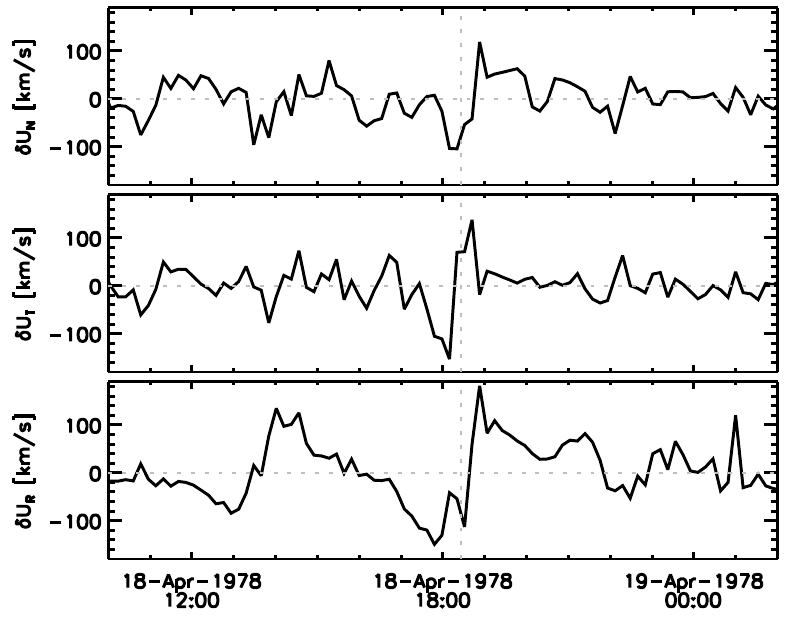}
\end{center}
\caption{Fluctuating proton flow velocity components as derived from the {\em Helios-2} observations.
The timestamp is in UT.}
\label{fig:u_obs_f}
\end{figure}

\begin{figure}
\begin{center}
\includegraphics[trim=0 0 0 0,width=8.0cm]{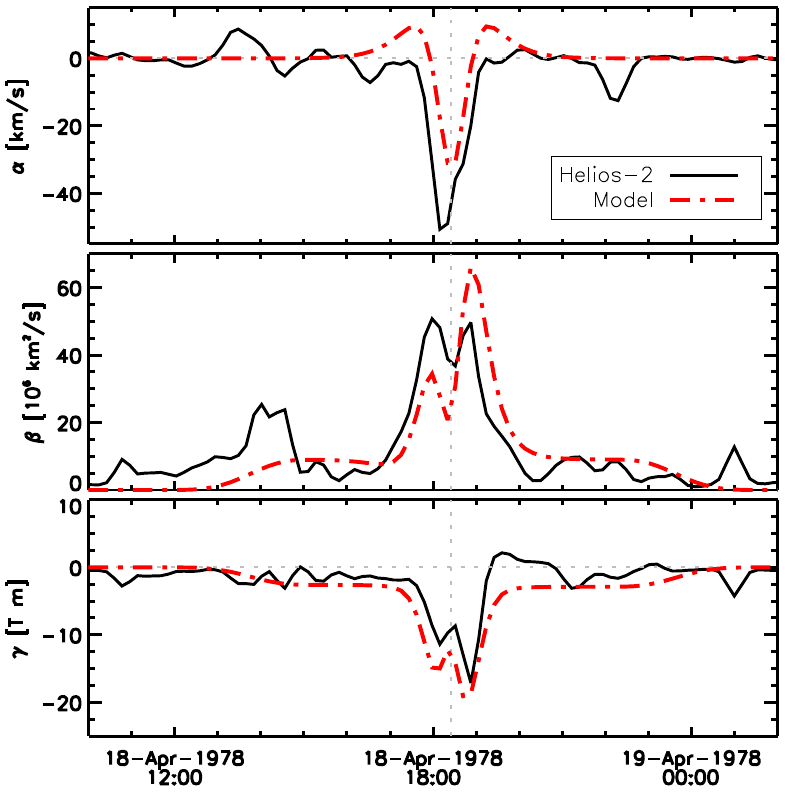}
\end{center}
\caption{Comparison of the $\alpha$, $\beta$ and $\gamma$ transport coefficient as derived from the {\em Helios-2} observations and the vortex model.
Plots have been smoothed with a Gaussian smoothing function ($\sigma_{\textrm{smooth}} = 0.9$).
The timestamp is in UT.}
\label{fig:alpha_beta_obs}
\label{fig:alpha_beta_1}
\end{figure}

\begin{figure}
\begin{center}
\includegraphics[trim=0 0 0 0,width=8.0cm]{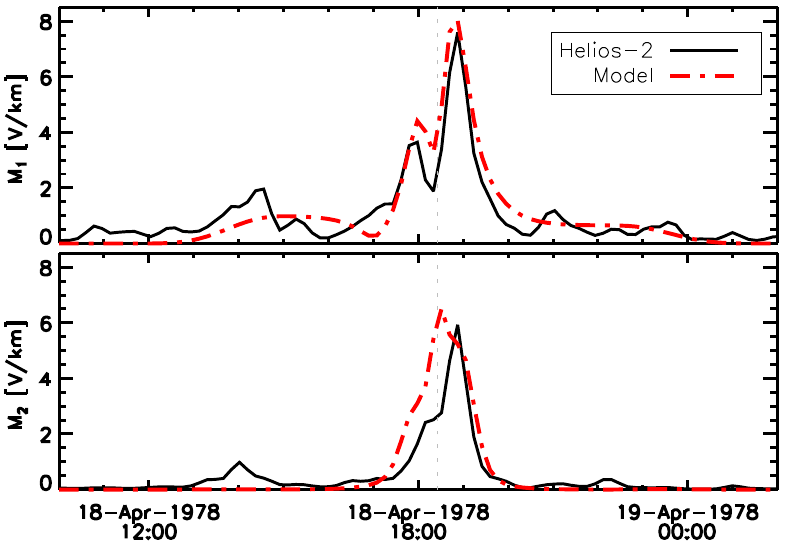}
\end{center}
\caption{Comparison of the magnitude of the electromotive force as derived from the {\em Helios-2} observations and the vortex model.
Upper panel: $M_1$; see Eq.\,(\ref{eq:em}).
Lower panel: $M_2$; see Eq.\,(\ref{eq:em_2}).
Plots have been smoothed with a Gaussian smoothing function ($\sigma_{\textrm{smooth}} = 0.9$).
The timestamp is in UT.}
\label{fig:em_comp_obs}
\end{figure}

\begin{figure}
\begin{center}
\includegraphics[trim=0 0 0 0,width=8.0cm]{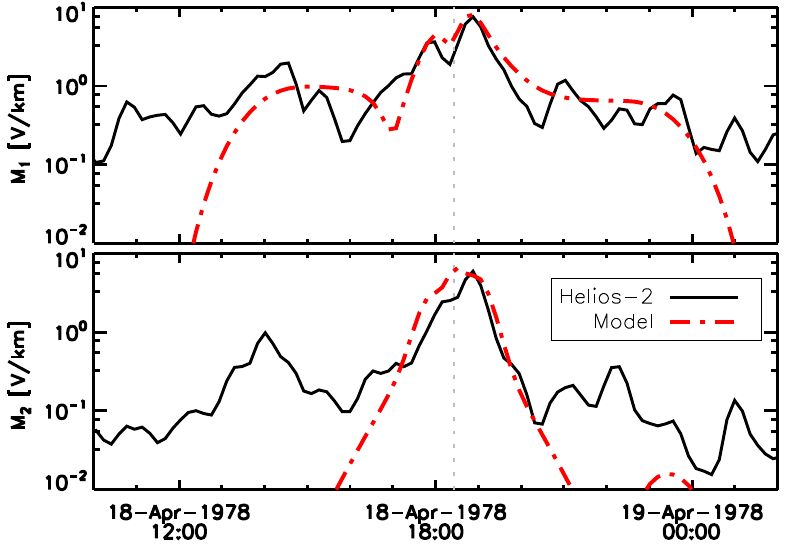}
\end{center}
\caption{Same as Fig.\,\ref{fig:em_comp_obs}, but with the magnitude of the electromotive force displayed on a logarithmic scale.}
\label{fig:em_comp_obs_log}
\end{figure}

Figs.\,\ref{fig:b_obs}--\ref{fig:em_comp_obs_log} show the results of the {\em Helios} data analysis.
The center of a structure in the magnetic transient that we interpret as a vortical structure is marked by a vertical dotted line.
Our model is in the coordinate system $x=N$, $y=T$, and $z=R$.
When comparing the magnetic field and plasma flow velocity components to our model, we have to keep in mind that we assume the case of a vortical structure that lies entirely in the $yz$-plane (or $TR$-plane in case of the observation).
Therefore, our model does not feature a radial profile in the $x(N)$ component.
Also, by assuming a one-dimensional cut through the vortex center, the radial magnetic field component does not show any helical features and is constant.

Figs.\,\ref{fig:b_obs} and\,\ref{fig:u_obs} show the observed RTN components and the magnitude of the magnetic field and plasma flow velocity respectively.
In comparison to our model (Fig.\,\ref{fig:b_model1} and \ref{fig:u_model1}), there is an initial increase and subsequent drop in all three magnetic field components, resulting in a change of sign around the center of the possible vortical structure, which could be explained by the structure not being restricted to the $TR$-plane.
The plasma flow velocity components show a similar behavior with a decrease in front of the structure followed by an increase after the structure.
However, the most prominent resemblance to the model can be found in the magnitude values with the magnitude of the plasma flow velocity resembling a shock front with a sudden increase followed by a slow relaxation over several hours.
Simultanously, the magnitude of the magnetic field features a temporary increase around the region of interest which could be caused by induction work on $\vec{B}$, indicating the possibility of vortical plasma motions.


Fig.\,\ref{fig:b_obs_m} shows the mean magnetic field components.
We observe that the normal and tangential components are not zero as they are in our model.
As in the model, the radial component dominates with a magnitude of about $30\,\rm{nT}$.
While the observed mean fields are not constant, they do not show any sudden changes either, which shows this assumption is reasonable on the time scale of the fluctuations.

When using a streamwise coordinate system, the normal and tangential mean plasma flow velocity components are supposed to be zero and indeed that is the picture we get in the upper two panels of Fig.\,\ref{fig:u_obs_m}.
As for the mean magnetic field, the mean plasma flow velocity is not a constant but the changes are slow enough for the mean vorticity $\vec{\Omega_0}$ to be assumed constant on the fluctuation time scale.

We are especially interested in the components of the fluctuations of the fields: as they are centered around zero, they are useful for identifying changes of sign that are indicators of vortical structures.
The fluctuating magnetic field components are shown in Fig.\,\ref{fig:b_obs_f}, where we see interesting behavior of all three components around the time of the presumed vortical structure.
The tangential and radial components feature a change of sign from positive to negative while the normal component $\delta B_N$ shows an isolated peak from about $+20$ to $-80\,\rm{nT}$ shortly after the center of the transition layer, at the same time the magnitude of the magnetic field reaches its maximum.
This can be interpreted as a helical structure that would explain the increase in the magnetic field magnitude as a result of induction work.

The fluctuations of the plasma flow velocity components (Fig.\,\ref{fig:u_obs_f}) interestingly change sign in all components before the $\delta B_N$ peak, and again after the event has passed the spacecraft.
After the event, the velocity components are more alike fluctuations around zero, similar to the radial profile we used in our model.
While this does not directly prove the existence of a vortex in the solar wind, such behavior would be expected from vortical plasma motions.
Additionally, we find that the cross-helicity effect is relevant for the turbulent transport within the solar wind.

The transport coefficients are calculated with Eq.\,(\ref{eq:alpha1beta1}) using numerical derivatives.
Just as in the model (see Fig\,\ref{fig:alpha_beta_1}), we find a single peak in the kinetic helicity ($\alpha$), as well as an asymmetric double peak in the turbulent diffusion ($\beta$) and the cross-helicity ($\gamma$) coefficient; see Fig.\,\ref{fig:alpha_beta_obs}.
However, the peak in $\alpha$ is higher than in the model and there are no minor peaks with opposite sign around it, which means that if the observed plasma flow velocity describes a vortex it must follow a different radial profile than assumed in our model.
In both observation and model, we find a double-peak structure in $\beta$, even though it is more symmetric in the observation.
As in the model, we see a plateau phase in $\beta$ and $\gamma$ around their peaks in the vortex center.
The $\gamma$ coefficient from the observation fits well in its asymmetry and amplitude to our model prediction.


In the upper panel of Fig.\,\ref{fig:em_comp_obs} we show the magnitude of the turbulent electromotive force $M_1$ (solid line) that shows a basic agreement with our model prediction (red dotted line).
Apparently, our model of helical magnetic field and vortical plasma motion fits reasonably well to the observed electromotive force formulation derived from mean-field electrodynamics ($M_1$).
We find a double peak at the vortex center with similar asymmetry and a spatially extended plateau phase.

For the absolute value of $M_2$ we find a peak next to the vortex center, similar to our model prediction; see the lower panel of Fig.\,\ref{fig:em_comp_obs}.
As in the model, there is no prominent plateau phase around the peak in $M_2$.
However, the overall agreement between model and observation is less good than for $M_1$: while the peak in our model prediction has a similar asymmetry and amplitude to the observed one, there is a slight spatial shift inbetween them.
This might stem from a too steep velocity gradient at our model's vortex center because the central peak is dominated by the cross-helicity term's contribution.

The increase to and the decrease from the plateau value can be better identified in a logarithmic plot; see Fig.\,\ref{fig:em_comp_obs_log}.

%%%%%%%%%%%%%%%%%%%%%%%%%%%%%%%%%%%%%%%%%%%%
\section{Discussion and outlook} \label{disc}

In order to understand the internal magnetical structure of CME shock fronts, we model vortical plasma flows and a helical magnetic field that may cause induction work and a turbulent electromotive force.
The {\em Helios} spacecraft are among the few that provided magnetic field and plasma flow velocity measurements within the inner heliosphere near 0.4\,\rm{au}.
Because these were single-spacecraft missions, we derive the kinetic helicity and turbulent diffusion transport coefficients, as well as the electromotive force, using Taylor's frozen-in flow hypothesis \citep{Taylor:1938}.
We reproduce the observed transport coefficients, $\alpha$, $\beta$, $\gamma$, and the magnitude of the electomotive force $M$ of a magnetic-transient event passing by the spacecraft with a model of a helical magnetic field and vortical plasma motion.
For the model and the observation alike, at the center of the transient event we find a single peak in the kinetic helicity ($\alpha$ coefficient) that is surrounded by a double peak in the turbulent diffussion ($\beta$ coefficient) and cross-helicity effect ($\gamma$ coefficient).
The magnitude of the electromotive force $M_1$ from mean-field electrodynamics has a clear double-peak structure, while the formulation $M_2$ that uses turbulent transport coefficients has a maximum peak at the vortex center of our model.
One possible reason for this disagreement might be that our model has a too strong velocity gradient at the vortex center.

Another explanation might be that we altered the original formulation of $M_2$ from theoretical and numerical dynamo studies for the application to in-situ measurements in the solar wind; see Sect.\,\ref{em-force}.
We use the Gaussian-convolution residual of the fields to calculate the vorticity and current density.
This makes our alternative formulation of the electromotive force $M_2$ conceptually different from the formalism used for numerical studies and theoretical models like \cite{Yoshizawa:1990}.
A better way of separating the background field and the fluctuation may be to refine the filtering procedure, in particular to further split the resolved GS from the unresolved SGS components in our fluctuations obtained with the Gaussian convolution; see Sec.\,\ref{background}.
However, direct measurements of the electromotive force in interplanetary space are rare \citep{Marsch+Tu:1992,Marsch+Tu:1993,Narita+Vörös:2018} and this work still proves the concept of an automatic CME detection method that could help planning future missions like {\em Solar Orbiter}.

With the inclusion of the cross-helicity term in $M_2$ we find that both formulations for the electromotive force $M_1$ and $M_2$ reproduce the observation from our model reasonably well.
The formulation for $M_2$ reflects a helically driven turbulent dynamo process in an incompressible medium.
For a CME shock front at 0.4\,\rm{au} this assumption might not sound applicable.
Nonetheless, the good agreement between model and observation would support the applicability of our alternative formulation.
Differences to the mean-field electrodynamics formulation $M_1$ still exist and could be caused, e.g., by the compressibility of the heliospheric plasma.

More sophisticated models of the turbulent transport coefficients, e.g. from mixing-length theory, might be better applicable for the calculation of the turbulent electromotive force in the solar wind.
It would be worth trying to see whether variations in our shock-front model would reproduce the observations better, e.g. with different spatial cuts not crossing the vortex center, a larger set of free parameters, or different background field definitions.
Regarding our choice of the background field direction, it is advisable to use non-parallel magnetic field and velocity vectors for distances from the Sun larger than 0.4\,\rm{au}.

The peak solar wind speed that we observe right after the shock front passes is about 300\,\rm{km/s} stronger than our vortex model suggests.
This gives room to improve our model parameters in future work, for example with a fitting procedure.

%%%%%%%%%%%%%%%%%%%%%%%%%%%%%%%%%%%%%%%%%%%%
\section{Conclusions} \label{conclusion}

Our work presents a method of estimating the kinetic and cross-helicity, as well as the turbulent diffusion from spacecraft observations of magnetic-transient events in the solar wind.
The method can be used as an indicator of possible vortical plasma motions and helical magnetic fields.
We suggest to use this method for further studies of the inner structure of CME shock fronts.

The magnetic transient event measured by {\em Helios-2} on 1978 April 18, 18:25:38 UT at 0.4\,\rm{au}, reveals a significant peak in both electromotive force formulations, $M_1$ and $M_2$.
The CME shock-front model that we construct is consistent with that event, if we assume a helical magnetic field and a vortical plasma flow, where the magnetic flux density rises from a background value of 25\,\rm{nT} to a plateau of about 35\,\rm{nT} already four hours before reaching a maximum of 80\,\rm{nT}, which lasts for about one hour as the shock passes by the spacecraft (Figs.\,\ref{fig:b_model1} and \ref{fig:b_obs}).
We find that the measured solar wind speeds are consistent with a vortical flow that departs from a slow background wind speed of about 450\,\rm{km/s} with an extended plateau phase also starting about four hours before the event.
Then, after a decrease of some 100\,\rm{km/s} in wind speed that lasts for up to one hour, there is a sharp increase to 600\,\rm{km/s} when the shock front passes by.
At that time, we observe a sign reversal in practically all velocity components after subtracting the background (Fig.\,\ref{fig:u_obs_f}), which supports our model assumption of vorticity, and hence co-existing magnetic helicity, in the solar wind.
This internal structure of the CME shock front produces a significant peak of 6--8\,\rm{V/km} in either formulation of the electromotive force (Fig.\,\ref{fig:em_comp_obs}).

As the method is designed for single-spacecraft measurements it can be used for the future {\em Solar Orbiter} mission that will initially have a similar orbit to that of the {\em Helios} spacecraft.
In comparison, {\em Solar Orbiter} will have the advantage of providing remote-sensing data in addition to in-situ measurements.
This will provide a supplementary method of identifying the arrival of a CME shock front.
Future work should include a statistical study of the transport coefficients and the electromotive force in multiple magnetic transients and CMEs in the inner heliosphere.

\vspace{2cm} % \newpage

%%%%%%%%%%%%%%%%%%%%%%%%%%%%%%%%%%%%%%%%%%%%
\acknowledgments
The authors would like to thank an anonymous referee explicitly for very helpful discussion.
B.H. implemented the data analysis and contributed to the manuscript text and figures.
This work is financially supported by the Austrian Space Applications Programme at the Austrian Research Promotion Agency, FFG ASAP-12 SOPHIE under contract 853994.
The observational data is provided by the {\em Helios} data archive in CDAWeb\footnote{\url{https://spdf.sci.gsfc.nasa.gov/pub/data/helios/}} located at the Space Physics Data Facility (NASA/GSFC).

\bibliography{Literatur}

\begin{thebibliography}{}
\expandafter\ifx\csname natexlab\endcsname\relax\def\natexlab#1{#1}\fi
\providecommand{\url}[1]{\href{#1}{#1}}

\bibitem[{{Balogh} {et~al.}(1997){Balogh}, {Dunlop}, {Cowley}, {Southwood},
  {Thomlinson}, {Glassmeier}, {Musmann}, {Luhr}, {Buchert}, {Acuna},
  {Fairfield}, {Slavin}, {Riedler}, {Schwingenschuh}, \&
  {Kivelson}}]{Balogh+al:1997}
{Balogh}, A., {Dunlop}, M.~W., {Cowley}, S.~W.~H., {et~al.} 1997,
  \hypersetup{urlcolor=magenta}\href{https://doi.org/10.1023/A:1004970907748}{Space~Sci.~Rev.},
   \hypersetup{urlcolor=blue}\href{http://adsabs.harvard.edu/abs/1997SSRv...79...65B}{79,
  65--91}

\bibitem[{{Bodin} \& {Newton}(1980)}]{Bodin+Newton:1980}
{Bodin}, H.~A.~B., \& {Newton}, A.~A. 1980, Nuclear Fusion,
  \hypersetup{urlcolor=blue}\href{http://adsabs.harvard.edu/abs/1980NucFu..20.1255B}{20,
  1255--1324}

\bibitem[{{Brandenburg}(2001)}]{Brandenburg:2001}
{Brandenburg}, A. 2001,
  \hypersetup{urlcolor=magenta}\href{https://doi.org/10.1086/319783}{ApJ},
  \hypersetup{urlcolor=blue}\href{http://adsabs.harvard.edu/abs/2001ApJ...550..824B}{550,
  824--840}

\bibitem[{{Brandenburg} \& {Subramanian}(2005)}]{Brandenburg+Subramanian:2005}
{Brandenburg}, A., \& {Subramanian}, K. 2005,
  \hypersetup{urlcolor=magenta}\href{https://doi.org/10.1016/j.physrep.2005.06.005}{Phys.~Rep.},
   \hypersetup{urlcolor=blue}\href{http://adsabs.harvard.edu/abs/2005PhR...417....1B}{417,
  1--209}

\bibitem[{{Bruno} \& {Carbone}(2013)}]{Bruno+Carbone:2013}
{Bruno}, R., \& {Carbone}, V. 2013,
  \hypersetup{urlcolor=magenta}\href{https://doi.org/10.12942/lrsp-2013-2}{Living
  Reviews in Solar Physics},
  \hypersetup{urlcolor=blue}\href{http://adsabs.harvard.edu/abs/2013LRSP...10....2B}{10,
  2}

\bibitem[{{Fr{\"a}nz} \& {Harper}(2002)}]{Fränz+Harper:2002}
{Fr{\"a}nz}, M., \& {Harper}, D. 2002,
  \hypersetup{urlcolor=magenta}\href{https://doi.org/10.1016/S0032-0633(01)00119-2}{Planet.~\&~Space~Sci.},
   \hypersetup{urlcolor=blue}\href{http://adsabs.harvard.edu/abs/2002P&SS...50..217F}{50,
  217--233}

\bibitem[{{Hamba}(1992)}]{Hamba:1992}
{Hamba}, F. 1992,
  \hypersetup{urlcolor=magenta}\href{https://doi.org/10.1063/1.858314}{Physics
  of Fluids},
  \hypersetup{urlcolor=blue}\href{http://adsabs.harvard.edu/abs/1992PhFl....4..441H}{4,
  441--450}

\bibitem[{{Ji} \& {Prager}(2002)}]{Ji+Prager:2002}
{Ji}, H., \& {Prager}, S.~C. 2002, Magnetohydrodynamics,
  \hypersetup{urlcolor=blue}\href{http://adsabs.harvard.edu/abs/2002MHD....38..191J}{38,
  191--210}

\bibitem[{{Krause} \& {R{\"a}dler}(1980)}]{Krause+Rädler:1980}
{Krause}, F., \& {R{\"a}dler}, K.-H. 1980, {Mean-field Magnetohydrodynamics and
  Dynamo Theory}
  (\hypersetup{urlcolor=magenta}\href{https://doi.org/10.1017/S0022112082211190}{Oxford,
  Pergamon Press, Ltd.})

\bibitem[{{Marsch} \& {Tu}(1992)}]{Marsch+Tu:1992}
{Marsch}, E., \& {Tu}, C.~Y. 1992, in Solar Wind Seven, ed. E.~{Marsch} \&
  R.~{Schwenn}, Proceedings of the 3rd COSPAR Colloquium
  (\hypersetup{urlcolor=magenta}\href{https://doi.org/10.1016/B978-0-08-042049-3.50105-8}{Pergamon,
  Amsterdam}),
  \hypersetup{urlcolor=blue}\href{http://adsabs.harvard.edu/abs/1992sws..coll..505M}{505--510}

\bibitem[{{Marsch} \& {Tu}(1993)}]{Marsch+Tu:1993}
{Marsch}, E., \& {Tu}, C.-Y. 1993, in 157th IAU Symposium, Vol. 157, The Cosmic
  Dynamo, ed. F.~{Krause}, K.~H. {R{\"a}dler}, \& G.~{R{\"u}diger}
  (\hypersetup{urlcolor=magenta}\href{https://doi.org/10.1017/S0074180900173863}{Cambridge
  University Press}),
  \hypersetup{urlcolor=blue}\href{http://adsabs.harvard.edu/abs/1993IAUS..157...51M}{51}

\bibitem[{{Musmann} {et~al.}(1975){Musmann}, {Neubauer}, {Maier}, \&
  {Lammers}}]{Musmann+al:1975}
{Musmann}, G., {Neubauer}, F., {Maier}, A., \& {Lammers}, E. 1975,
  Raumfahrtforschung,
  \hypersetup{urlcolor=blue}\href{http://adsabs.harvard.edu/abs/1975RF.....19..232M}{19,
  232--237}

\bibitem[{{Narita} \& {V{\"o}r{\"o}s}(2018)}]{Narita+Vörös:2018}
{Narita}, Y., \& {V{\"o}r{\"o}s}, Z. 2018,
  \hypersetup{urlcolor=magenta}\href{https://doi.org/10.5194/angeo-36-101-2018}{An.~Geo.},
   \hypersetup{urlcolor=blue}\href{http://adsabs.harvard.edu/abs/2018AnGeo..36..101N}{36,
  101--106}

\bibitem[{{Pouquet} {et~al.}(1976){Pouquet}, {Frisch}, \&
  {Leorat}}]{Pouquet+al:1976}
{Pouquet}, A., {Frisch}, U., \& {Leorat}, J. 1976,
  \hypersetup{urlcolor=magenta}\href{https://doi.org/10.1017/S0022112076002140}{Journal
  of Fluid Mechanics},
  \hypersetup{urlcolor=blue}\href{http://adsabs.harvard.edu/abs/1976JFM....77..321P}{77,
  321--354}

\bibitem[{{Priest}(1982)}]{Priest:1982}
{Priest}, E.~R. 1982, {Solar Magnetohydrodynamics}, Boston Studies in the
  Philosophy of Science
  (\hypersetup{urlcolor=magenta}\href{https://doi.org/10.1007/978-94-009-7958-1}{D.
  Reidel, Dordrecht})

\bibitem[{{Schwenn} {et~al.}(1975){Schwenn}, {Rosenbauer}, \&
  {Miggenrieder}}]{Schwenn+al:1975}
{Schwenn}, R., {Rosenbauer}, H., \& {Miggenrieder}, H. 1975,
  Raumfahrtforschung,
  \hypersetup{urlcolor=blue}\href{http://adsabs.harvard.edu/abs/1975RF.....19..226S}{19,
  226--232}

\bibitem[{{Steenbeck} {et~al.}(1966){Steenbeck}, {Krause}, \&
  {R{\"a}dler}}]{Steenbeck+al:1966}
{Steenbeck}, M., {Krause}, F., \& {R{\"a}dler}, K.-H. 1966,
  \hypersetup{urlcolor=magenta}\href{https://doi.org/10.1515/zna-1966-0401}{Zeitschrift
  Naturforschung Teil A},
  \hypersetup{urlcolor=blue}\href{http://adsabs.harvard.edu/abs/1966ZNatA..21..369S}{21,
  369}

\bibitem[{{Sur} \& {Brandenburg}(2009)}]{Sur+Brandenburg:2009}
{Sur}, S., \& {Brandenburg}, A. 2009,
  \hypersetup{urlcolor=magenta}\href{https://doi.org/10.1111/j.1365-2966.2009.15254.x}{MNRAS},
   \hypersetup{urlcolor=blue}\href{http://adsabs.harvard.edu/abs/2009MNRAS.399..273S}{399,
  273--280}

\bibitem[{{Taylor}(1938)}]{Taylor:1938}
{Taylor}, G.~I. 1938,
  \hypersetup{urlcolor=magenta}\href{https://doi.org/10.1098/rspa.1938.0032}{Proceedings
  of the Royal Society of London Series A},
  \hypersetup{urlcolor=blue}\href{http://adsabs.harvard.edu/abs/1938RSPSA.164..476T}{164,
  476--490}

\bibitem[{{Torbert} {et~al.}(2016){Torbert}, {Russell}, {Magnes}, {Ergun},
  {Lindqvist}, {LeContel}, {Vaith}, {Macri}, {Myers}, {Rau}, {Needell}, {King},
  {Granoff}, {Chutter}, {Dors}, {Olsson}, {Khotyaintsev}, {Eriksson},
  {Kletzing}, {Bounds}, {Anderson}, {Baumjohann}, {Steller}, {Bromund}, {Le},
  {Nakamura}, {Strangeway}, {Leinweber}, {Tucker}, {Westfall}, {Fischer},
  {Plaschke}, {Porter}, \& {Lappalainen}}]{Torbert+al:2016}
{Torbert}, R.~B., {Russell}, C.~T., {Magnes}, W., {et~al.} 2016,
  \hypersetup{urlcolor=magenta}\href{https://doi.org/10.1007/s11214-014-0109-8}{Space~Sci.~Rev.},
   \hypersetup{urlcolor=blue}\href{http://adsabs.harvard.edu/abs/2016SSRv..199..105T}{199,
  105--135}

\bibitem[{{Yokoi}(2013)}]{Yokoi:2013}
{Yokoi}, N. 2013,
  \hypersetup{urlcolor=magenta}\href{https://doi.org/10.1080/03091929.2012.754022}{Geophysical
  and Astrophysical Fluid Dynamics},
  \hypersetup{urlcolor=blue}\href{http://adsabs.harvard.edu/abs/2013GApFD.107..114Y}{107,
  114--184}

\bibitem[{{Yoshizawa}(1990)}]{Yoshizawa:1990}
{Yoshizawa}, A. 1990,
  \hypersetup{urlcolor=magenta}\href{https://doi.org/10.1063/1.859484}{Physics
  of Fluids B},
  \hypersetup{urlcolor=blue}\href{http://adsabs.harvard.edu/abs/1990PhFlB...2.1589Y}{2,
  1589--1600}

\end{thebibliography}
\bibliographystyle{aasjournal}

\end{document}